\documentclass{emulateapj}

\shorttitle{Spectrum of cosmic rays, produced in supernova remnants}
\shortauthors{E.G. Berezhko \& H.J.V\"olk}

\begin{document}

\title{Spectrum of cosmic rays, produced in supernova remnants}

\author{E.G.~Berezhko\altaffilmark{1}
        H.J.V\"olk\altaffilmark{2}
}
\altaffiltext{1}{Yu.G. Shafer Institute of Cosmophysical Research and Aeronomy,
                     31 Lenin Ave., 677980 Yakutsk, Russia}
		     
\altaffiltext{2}{Max Planck Institut f\"ur Kernphysik,
                Postfach 103980, D-69029 Heidelberg, Germany}

\email{berezhko@ikfia.ysn.ru, Heinrich.Voelk@mpi-hd.mpg.de}

\begin{abstract} 
     Nonlinear kinetic theory of cosmic ray (CR) acceleration in supernova
     remnants is employed to calculate CR spectra. The magnetic field in SNRs
     is assumed to be significantly amplified by the efficiently accelerating
     nuclear CR component.  It is shown that the calculated CR spectra agree in
     a satisfactory way with the existing measurements up to the energy
     $10^{17}$~eV.  The power law spectrum of protons extends up to the energy
     $3\times 10^{15}$~eV with a subsequent exponential cutoff. It gives a
     natural explanation for the observed knee in the Galactic CR spectrum.
     The maximum energy of the accelerated nuclei is proportional to their
     charge number $Z$. Therefore the break in the Galactic CR spectrum is the
     result of the contribution of progressively heavier species in the overall
     CR spectrum so that at $10^{17}$~eV the CR spectrum is dominated by iron
     group nuclei.  It is shown that this component plus a suitably chosen
     extragalactic CR component can give a consistent description for the
     entire Galactic CR spectrum.
\end{abstract}
 


\section{Introduction}
The main reason why supernova remnants (SNRs) are usually considered as the
cosmic ray (CR) source is a simple argument about the energy required to
sustain the Galactic cosmic ray (GCR) population against loss by escape,
nuclear interactions and ionization energy loss.  Supernovae have enough power
to drive the GCR acceleration if there exists a mechanism for channeling about
10\% of the mechanical energy release into relativistic particles. The high
velocity ejecta produced in the supernova (SN) explosion interacts with the
ambient medium to produce a strong blast wave.  This outer shock may accelerate
a small suprathermal fraction of the ambient plasma to high energies.

The only theory of particle acceleration which at present is sufficiently well
developed and specific to allow quantitative model calculations is diffusive
acceleration applied to the strong outer shock associated with SNRs
\citep[e.g. see][for review]{ber05}. Considerable efforts have been made during
the last years to empirically confirm the theoretical expectation that the main
part of GCRs indeed originates in SNRs. Theoretically, progress in the solution
of this problem has been due to the development of a kinetic nonlinear theory
of diffusive shock acceleration \citep{byk96,bv97,bv00}. This theory attempts
to include all the most relevant physical factors, essential for the evolution
and CR acceleration in a SNR, at least in its early very energetic stages, and
it is able to make quantitative predictions of the expected properties of CRs
produced in young SNRs and their nonthermal radiation. The application of the
theory to individual SNRs and their known synchrotron emission \citep{vlk03,
ber05,bv06} has demonstrated its capability of explaining the observed SNR
properties and in calculating new effects like the extent of magnetic field
amplification which leads to the concentration of the highest-energy electrons
in a very thin shell just behind the shock. The theory should therefore be able
to explain major characteristics of the observed GCR spectrum up to an energy
of $\sim 10^{15}$~eV \citep{bk99} under the assumption that the ambient
magnetic field strength considerably exceeds typical ISM values.

Here we apply kinetic nonlinear theory assuming a time-dependent amplified
magnetic field, consistent with multi-wavelength evidence from individual
objects, in order to approximately calculate the spectra of CRs produced in
Galactic SNRs. It is shown that these spectra are approximately consistent with
the existing measurements of GCR spectra up to an energy of $10^{17}$~eV.

\section{Model}
Our nonlinear model is based on a fully time-dependent solution of the CR
transport equation together with the gas dynamic equations in spherical
symmetry. Since all relevant equations, initial and boundary conditions for
this model have already been described in detail elsewhere
\citep[e.g.][]{bv04}, we do not present them here and only briefly discuss
the most important aspects below.

The basic scenario is that of a point explosion, where the supernova explosion
ejects an expanding shell of matter with total energy $E_{\mathrm {sn}}$ and
mass $M_{\mathrm {ej}}$ into the surrounding ISM.  Due to the streaming
instability CRs efficiently excite large-amplitude magnetic fluctuations
upstream of the SN shock \citep[e.g.][]{bell78}. Since these fluctuations
scatter CRs extremely strongly, we assume the CR diffusion coefficient to be as
small as the Bohm limit
$\kappa (p)=pvc/(3ZeB)$, 
where $e$ and $m$ are the proton charge and mass, $v$ and $p$ denote the
particle velocity and momentum, $Z$ is the particle charge number, $B$ is the
magnetic field strength, and $c$ is the speed of light. Note that due to the
dependence $\kappa \propto vp/Z$ the shock-produced ion distributions depend on
rigidity
with the cutoff energy $\epsilon_{\mathrm {max}} \propto
Z$~\citep[e.g.][]{byk96}.

If $B_\mathrm{ISM}$ is the preexisting field in the surrounding ISM, then the Bohm
limit implies that the instability growth is restricted by some nonlinear
mechanism to the level $\delta B\sim B_{\mathrm {ISM}}$, where $\delta B$ is
the wave field. An earlier attempt for a nonlinear description of the magnetic
field evolution in a numerical model \citep{lb00,bl01} and more recent
instability calculations \citep {bell04} concluded that a considerable
amplification to what we call an effective magnetic field $B_0\gg B_{\mathrm
{ISM}}$ should occur. The Bohm limit is then expected to refer to this
amplified field. We adopt this point of view here.

From an analysis of the synchrotron spectrum of SN~1006, Cassiopeia~A, Tycho's
SNR \citep[see][for review]{ber05} such a strong magnetic field amplification
can only be produced as a nonlinear effect by a very efficiently accelerated
nuclear CR component.  The same large effective magnetic field is required by
the comparison of our selfconsistent theory with the morphology of the observed
X-ray synchrotron emission, in particular, its spatial fine structure.

In fact, for all the thoroughly
studied young SNRs, the ratio of magnetic field energy density $B_0^2/8\pi$ in
the upstream region of the shock precursor to the CR pressure $P_c$ is about
the same \citep{vbk05}:
$B_0^2/(8\pi P_{\mathrm {c}}) \approx 5\times 10^{-3}$.
We note that this amplified magnetic field $B_0$ exceeds the typical ISM
value $B_{\mathrm{ISM}}\approx 5$~$\mu$G during much of
the early evolution. Our present calculations assume that this is the
case for most of the evolution of the SNR and to this extent
the results of our calculations below are insensitive to the concrete value of
$B_{\mathrm {ISM}}$ except in the very late SNR evolutionary phase. 

The number of suprathermal protons injected into the acceleration process is
described by a dimensionless injection parameter $\eta$ which is a fixed
fraction of the number of ISM particles entering the shock front.  We adopt
here a value $\eta\sim 10^{-4}$, which is consistent with the theoretical
expectation \citep{vbk03} and which is close to the values determined
individually for the three objects SN~1006, Tycho's SNR, and Cas~A.

The injection rate of ions heavier than protons can not be calculated
with the required precision. Therefore we use ion injection rates which provide
ion-to-proton ratios as observed in the GCRs at an energy of $~1$~TeV. The
physical factors (ion injection rate and acceleration efficiency) which
determine this ratio were discussed by \citet{bk99}.

The overall CR spectrum $N(\epsilon_k,T_\mathrm{SN})$ 
is formed during the active period
of SNR evolution which lasts up to the time 
$T_\mathrm{SN}$ when the SN shock becomes
too weak to accelerate efficiently a new portion of freshly injected particles.
After their release from the parent SNRs the accelerated CRs occupy the
confinement volume more or less uniformly with an intensity
$J(\epsilon_k)\propto \tau_{\mathrm{esc}}(R)N(\epsilon_k)$
where $\tau_{\mathrm{esc}}(R)$ is the mean residence time, $R=pc/(Ze)$ is
rigidity, $p$ is momentum, $\epsilon =Amc^2+\epsilon_k$ and $\epsilon_k$ are
total and kinetic particle energy, respectively.

\section{Results and Discussion}

We use the values $E_\mathrm{SN}=10^{51}$~erg for the explosion energy and
$M_\mathrm{ej}=1.4 M_{\odot}$ for the ejecta mass which are typical for SNe~Ia
in a uniform ISM. Note that the main fraction of the core collapse SNe has
relatively small initial progenitor star masses between 8 and 15 $M_{\odot}$
which therefore do not significantly modify the surrounding ISM through the
main sequence wind of the progenitor star \citep[e.g.][]{abb}. SNR evolution in
this case is similar to that of SNe~Ia.

The active phase of the average SNR as CR source was
assumed to last until an age of $T_{\mathrm{SN}}\approx 10^5$~yr \citep{sp03}. 
\begin{figure}
\plotone{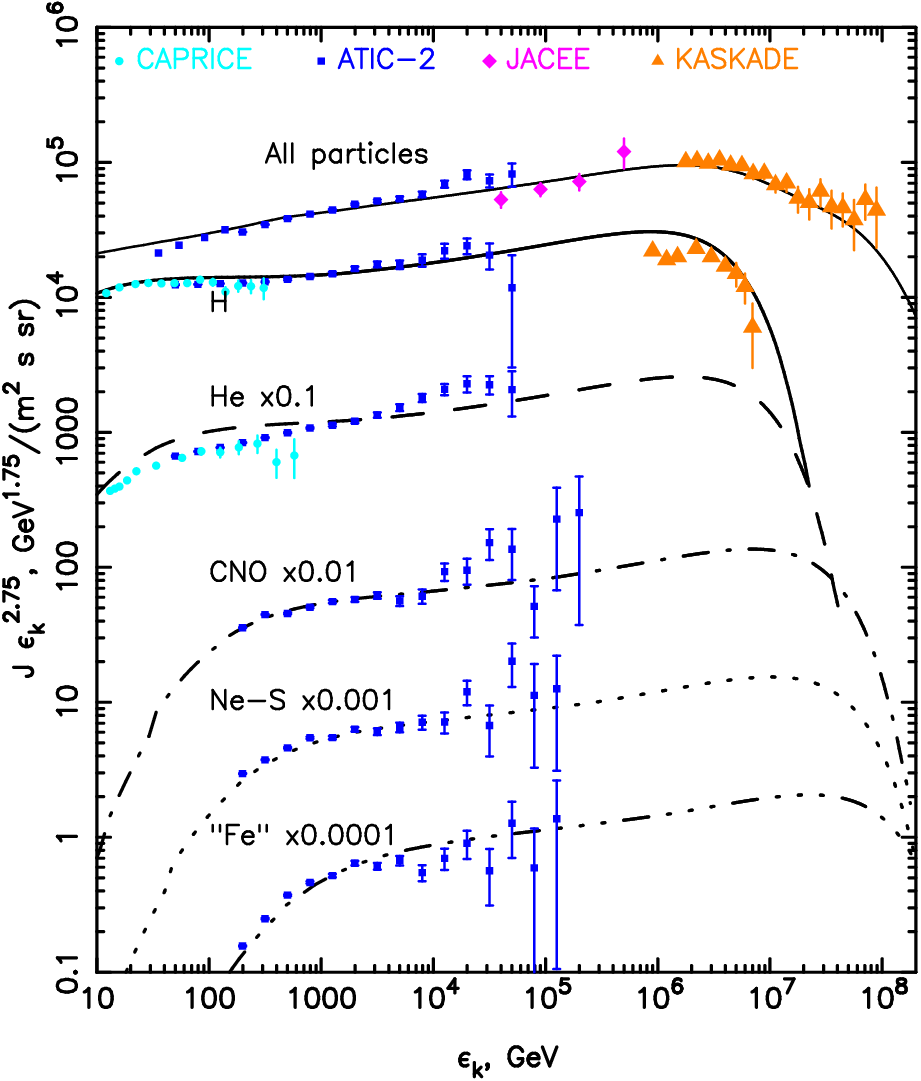}
\caption{GCR intensities at the Solar system as a function of particle kinetic
energy. Curves represent the CR spectra calculated for a proton injection rate
$\eta=10^{-4}$, ion-to-proton ratios which provide a fit for the data at
$\epsilon_k=1$~TeV, and for $\mu=0.75$.  Experimental data obtained in the
CAPRICE \citep{caprice}, ATIC-2 \citep{atic2}, JACEE \citep{jacee} and KASCADE
\citep{kascade} experiments are shown as well.}
\label{f1}
\end{figure}

In Fig.1 we present the calculated intensities $J(\epsilon_k)\times
\epsilon_k^{2.75}$ of protons (H), Helium, three groups of heavier nuclei, and
"All particles" as a function of kinetic energy, as solutions of the
nonlinear equations. Here we have used
$\tau_{\mathrm{esc}} \propto R^{-\mu}$, with $\mu = 0.75$. The results of the
recent experiments CAPRICE, ATIC-2 , JACEE and KASCADE, which in our view are
the most reliable ones, agree quite well with this theoretical calculation up
to the energy $\epsilon_k\approx 10^{17}$~eV. One can see that the theory fits
the existing data in a satisfactory way up to the energy $\epsilon_k \approx
10^{17}$~eV. The main exception -- and difficulty -- is the Helium spectrum as
measured in the recent ATIC-2 balloon experiment which is noticeably harder
than the proton spectrum, in contrast to the theoretical expectation.

The second difficulty for the present theory are the calculated very hard {\it
overall} CR source spectra which, say for protons, have the form $N\propto
\epsilon_k^{-1.9}$ for $\epsilon_k<10^{15}$~eV.  This spectrum is noticably
harder than the source spectrum $N\propto \epsilon_k^{-2.1}$, deduced in the
framework of our preferred CR propagation model which includes a selfconsistent
halo within a Galactic wind \citep{ptuskin97}. Note that this latter model
explains well the existing data, except for the observed low GCR anisotropy
which is at the moment a common problem for all CR diffusion models without
re-acceleration \citep[e.g.][]{ptuskin06}, even though the anisotropy is
presumably determined by the local structure of the Galactic magnetic field and
may deviate significantly from the global characteristics of the propagation
model \citep{ptuskin97}. However, \citet{pz03,pz05} have argued that already in
the middle Sedov phase nonlinear dissipation will increasingly reduce the field
amplification and particle scattering below the Bohm limit. High-energy
particles will then increasingly leave the SNR which can only continue to
accelerate lower and lower energy particles. Although there is a need for a
more detailed analysis of this effect, it 
should lead to some softening of
the overall GCR spectrum as the result of escape at high and continued
acceleration at lower energies in older SNRs. In this sense we do not consider
the calculated overall source spectra a problem for the arguments put forward
here. We will study this question in detail in a subsequent paper.

According to Fig.1 the knee in the observed all-particle GCR spectrum has to be
attributed to the maximum energy of protons, produced in SNRs. The steepening
of the all particle GCR spectrum above the knee energy $3\times 10^{15}$~eV is
a result of the progressively decreasing contribution of light CR nuclei with
increasing energy. Such a scenario is confirmed by the KASCADE experiment which
shows relatively sharp cutoffs of the spectra of various GCR species at
energies $\epsilon_{\mathrm{max}}\approx 3Z\times10^{15}$~eV \citep{kascade},
so that at energy $\epsilon_k \sim 10^{17}$~eV the GCR spectrum is expected to be
dominated by the contribution from the iron nuclei.

We note that the maximum CR energy in the overall CR spectrum is determined by
the CRs produced in the very beginning of the Sedov phase.  During the free
expansion phase the SN shock produces even higher-energy particles as a result
of the higher shock velocities and magnetic field strengths.  However, since
the mass in the very fast ejecta that produce them is very small, these
particles contribute only a very steep part (tail) of the final overall CR
spectrum, as was shown earlier \citep{bv04} for type Ia SNe. For this general
reason the same should also be true for wind SNe. As a result these particles
play no role for the population of Galactic CRs -- in contrast to the
assumption of \citet{bl01} which has been followed by \citet{hillas06}.
\begin{figure}
\plotone{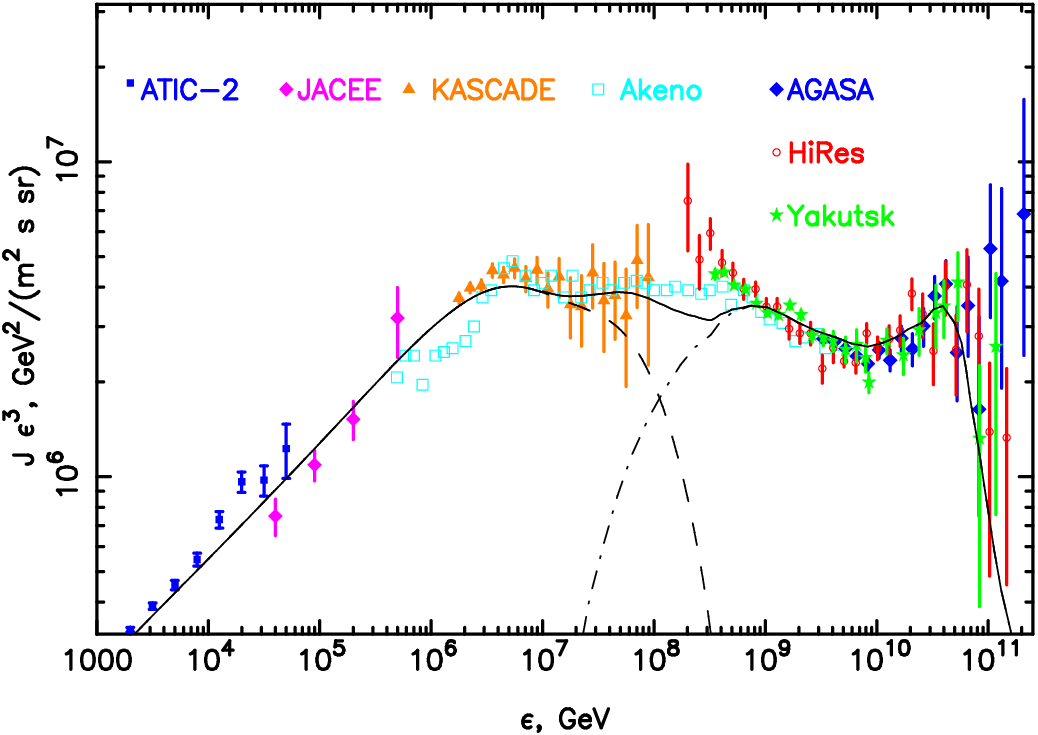}
\caption{All-particle GCR intensity as a function of total particle
energy. The dashed line
represents the Galactic component, which is the all-particle spectrum from
Fig.1. The dash-dotted line represents the assumed extragalactic
component. Experimental data obtained in the ATIC-2, JACEE, KASCADE, Akeno -
AGASA \citep{takeda03}, HiRes \citep{abbasi05} and Yakutsk \citep{egor04}
experiments are shown as well.}
\label{f2}
\end{figure}

In Fig.2 we present an all-particle spectrum which includes two components: (i)
CRs produced in SNRs and (ii) extragalactic CRs, protons plus 10\% Helium,
presumably produced in Active Galactic Nuclei \citep{aloisio06}. The second
component has been chosen to have a power-law source spectrum
$J_{\mathrm{s}}\propto \epsilon^{-2.7}$ above an energy $\epsilon\sim
10^{16}$~eV up to energies in excess of $10^{20}$~eV.

Compared with the source spectrum $J_{\mathrm{s}}(\epsilon)$, the component
$J(\epsilon)$, observed in the Galaxy, is modified by two factors. At energies
$\epsilon>10^{18}$~eV the shape of $J(\epsilon)$ is influenced by the energy
losses of CRs in intergalactic space as a result of their interaction with the
cosmic microwave background that leads to the formation of a "dip" structure at
$\epsilon\sim 10^{19}$~eV, and to a GZK-cutoff for $\epsilon>3\times
10^{19}$~eV \citep{aloisio06}.

For $\epsilon<10^{18}$~eV the spectrum $J(\epsilon)$ is determined by the
character of CR propagation in intergalactic space.  Since we assume the
existence of a Galactic Wind, CRs penetrating into the Galaxy from outside are
in addition subject to modulation by the wind. We describe this effect by the
modulation factor $f=\exp (-\epsilon_{\mathrm{m}}/\epsilon)$, where the maximum
CR energy modulated by the Galactic Wind is about
$\epsilon_{\mathrm{m}}=10^{17}$~eV \citep{vz04}.

Cf. \citet{aloisio06}, presenting  data of the AGASA, 
Yakutsk and HiRes detectors in Fig.2, we shift
the energies by a factor of $\lambda=0.9$, 0.85 and 1.2 respectively, so that
the measured CR fluxes agree with each other.

According to Fig.2 the calculated GCR spectrum is in reasonable agreement with
the existing data. It leads us to the following conclusion: if the observed CR
spectrum at energies $\epsilon>10^{18}$~eV is indeed dominated by the
contribution from extragalactic sources [the so-called "dip-model" of
\citet{aloisio06}], then we do not need any other Galactic source population
except SNRs, as calculated above.  However, if the extragalactic sources
produce in reality a much harder spectrum $J_{\mathrm{s}}\propto \epsilon^{-2}$
(the so-called "ankle-model") then their contribution becomes dominant only at
energies $\epsilon>10^{19}$~eV. Therefore, to fit the observed GCR spectrum an
additional Galactic source population is required whose contribution is
essential in the energy range $10^{17}<\epsilon<10^{19}$~eV \citep{hillas06}. 
It could possibly
result from CR reacceleration processes \citep[e.g.][]{bk99}, for example, in
the interaction regions of the Galactic Wind induced by the spiral structure in
the Galactic Disk \citep{vz04}. For further Galactic particle sources,
see \citet{hoerandel07}.
\begin{figure}
\plotone{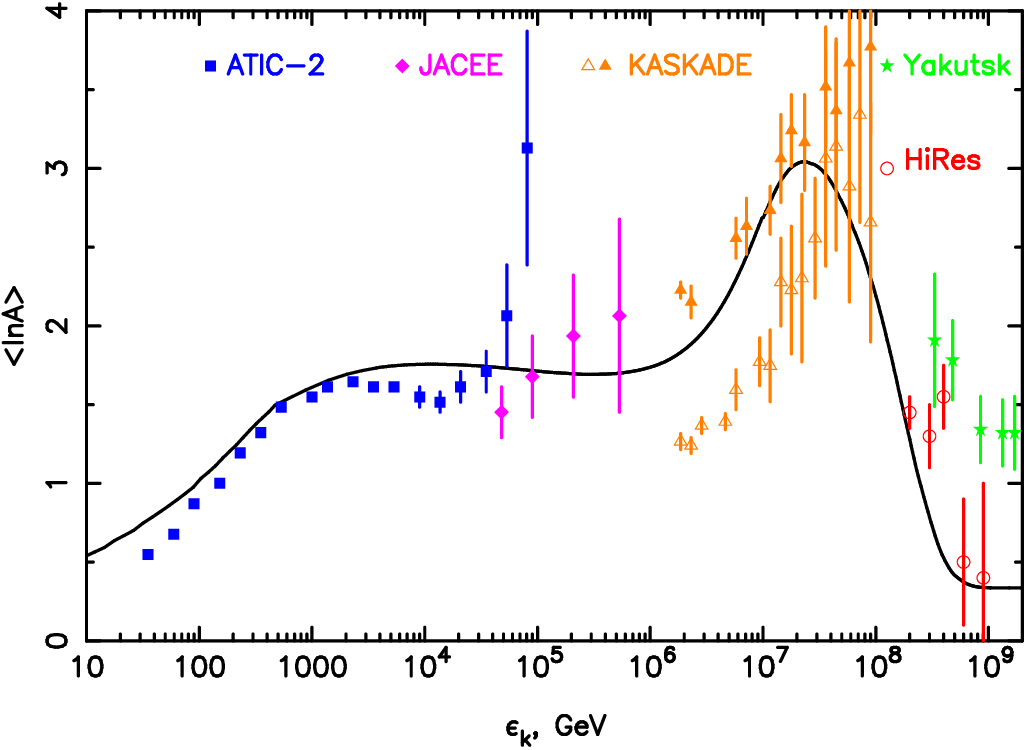}
\caption{Mean logarithm of the CR nucleus atomic number as a function of
  particle kinetic energy.  Experimental data obtained in the ATIC-2, JACEE,
  KASCADE \citep[QGSJET and SYBYLL;][]{hoerandel05}, as well as in the HiRes
  \citep[QGSJET;][]{hoerandel03} and Yakutsk \citep{ivan03} experiments are
  shown.}
\label{f3}
\end{figure}

We note that these two scenarios for the extragalactic component predict a very
different CR chemical composition above about $3\times 10^{17}$~eV. Since
within the range $3Z\times 10^{15}<\epsilon<Z\times 10^{17}$~eV reacceleration
produces a power law tail of the CR spectrum originally produced in SNRs, it is
clear that the observed CR spectrum is expected to be dominated by the iron
contribution at energies $10^{17}<\epsilon<10^{19}$~eV. It is very much
different from what is expected within the dip-model.

In order to illustrate the CR chemical composition, expected in the latter
case, we present in Fig.3 the mean logarithm of the GCR atomic number $A$.  At
energies $\epsilon_k<1$~TeV $A(\epsilon)$ increases with energy and for
$10^3<\epsilon_k<10^{6}$~GeV we have $<\ln A>~\approx 6$. Due to the dependence
of the maximum energy of CRs produced in SNRs, $\epsilon_{\mathrm{max}}\propto
Z$, on the charge number $Z$ and therefore on $A$, CRs become progressively
heavier above the proton cutoff energy $\epsilon_k\approx 3\times
10^{15}$~eV. The mean atomic number of the CRs produced in SNRs goes towards
the value $A=56$ as the energy approaches $10^{17}$~eV. However, already at
$\epsilon_k>10^{16}$~eV the contribution of extragalactic CRs becomes essential.
Therefore $<\ln A>$ reaches its peak value at $\epsilon_k\approx 2\times
10^{16}$~eV and then diminishes with increasing energy to the value $0.1 \times
\ln 4$.

The calculated atomic number agrees reasonably well with the existing data up
to a particle energy of $\epsilon_k\sim 10^{16}$~eV. At higher energies
$A(\epsilon_k)$ goes down with energy similar to the Yakutsk and HiRes data.
Quantitatively the calculated value of $<\ln A>$ agrees well with the HiRes
data.  Since the existing data demonstrate that above $10^{17}$~eV CRs become
lighter with energy, this could be considered to favor the dip-scenario.

\section{Summary}
Magnetic field amplification leads to a considerable increase of the maximum
energy of the CRs accelerated in SNRs. Calculations performed within nonlinear
kinetic theory demonstrate that the expected GCR spectrum, produced in SNRs,
fits the existing GCR data in a satisfactory way up to the energy
$10^{17}$~eV. The first knee in the observed all-particle GCR spectrum is
attributed to the maximum energy of protons that are produced in SNRs. The
steepening of the all-particle GCR spectrum above the knee energy $3\times
10^{15}$~eV is then the result of a progressive depression of the contribution
of light CR nuclei with increasing energy. Such a scenario is confirmed by the
KASCADE experiment which shows relatively sharp cutoff spectra of GCR species
at energies $\epsilon_{\mathrm{max}}\approx 3Z\times10^{15}$~eV
\citep{kascade}. A difficulty of the present computation are the resulting very
hard overall source spectra from SNRs. We expect however that inclusion of wave
dissipation and particle escape from SNRs at later phases of their evolution
will rectify this extreme.

If the Extragalactic CR component has a spectrum $J_{\mathrm{s}}\propto
\epsilon^{-2.7}$ so that it dominates the observed GCR spectrum already at
$\epsilon\geq 10^{18}$~eV, then GCRs at lower energies are produced in SNRs
without significant contribution from other Galactic CR sources.  The GCR
spectrum up to the energy $\epsilon=10^{17}$~eV is dominated by the
contribution of SNRs, whereas at $\epsilon>10^{18}$~eV GCRs are predominantly
Extragalactic. If the Extragalactic source spectrum is much harder,
$J_{\mathrm{s}}\propto \epsilon^{-2}$, then the transition from the Galactic to
the Extragalactic component is expected at higher energy $\epsilon\sim
10^{19}$~eV.  Since the expected GCR chemical composition at
$10^{17}-10^{18}$~eV is very different in these two cases, an experimental
study of GCR composition could discriminate them. The existing HiRes and
Yakutsk data favor the first, so-called dip scenario.

\acknowledgements
EGB acknowledges the hospitality of the Max-Plank-Institut f\"ur Kernphysik
where part of this work was carried out. This work has been supported in
part by the Russian Foundation for Basic Research (grant 07-02-00221).


\begin{thebibliography}{}

\bibitem[Abbasi et al.(2005)]{abbasi05}
Abbasi, R. U. et al. 2005, astro-ph/0501317

\bibitem[Abbott(1982)]{abb}
Abbott, D.C. 1982, \apj, 263, 723

\bibitem[Aloisio et al.(2006)]{aloisio06}
Aloisio, R. et al. 2007, Astropart. Phys., 27, 76

\bibitem[Antoni et al.(2005)]{kascade}
Antoni, T. et al. 2005, Astropart. Phys., 24, 1

\bibitem[Asakimori et al.(2003)]{jacee}
Asakimori, K. et al. 2003, \apj, 502, 278

\bibitem[Bell(1978)]{bell78}
Bell, A.R. 1978, \mnras, 182, 147

\bibitem[Bell(2004)]{bell04}
Bell, A.R. 2004, \mnras, 353, 550

\bibitem[Bell \& Lucek (2001)]{bl01}
Bell, A.R. \& Lucek, S.G. 2001, \mnras, 321, 433

\bibitem[Berezhko et al.(1996)]{byk96}
Berezhko, E. G., Elshin, V. K., \& Ksenofontov, L. T. 1996, JETP, 82, 1

\bibitem[Berezhko \& V\"olk(1997)]{bv97}
Berezhko, E.G. \& V\"olk, H.J. 1997, Astropart. Phys. 7, 183

\bibitem[Berezhko \& V\"olk(2000)]{bv00}
Berezhko, E.G. \& V\"olk, H.J. 2000, A\&A, 357, 183

\bibitem[Berezhko \& Ksenofontov(1999)]{bk99}
Berezhko, E.G. \& Ksenofontov, L.T. 1999, JETPh 89, 391

\bibitem[Berezhko et al.(2003)]{sp03}
Berezhko, E.G. et al. 2003, \aap, 410, 189 

\bibitem[Berezhko \& V\"olk(2004)]{bv04}
Berezhko, E.G. \& V\"olk, H.J. 2004, \aap, 427, 525

\bibitem[Berezhko(2005)]{ber05}
Berezhko, E.G. 2005, Adv. Space Res., 35, 1031

\bibitem[Berezhko \& V\"olk (2006)]{bv06}
Berezhko, E. G. \& V\"olk, H. J. 2006, \aap, 451, 981

\bibitem[Boezio et al.(2003)]{caprice}
Boezio, M. et al. 2003, Astropart. Phys., 19, 583

\bibitem[Egorova et al.(2004)]{egor04}
Egorova, V.P. et al. 2004, Nucl. Phys. B (Proc. Suppl.), 136, 3

\bibitem[Ivanov et al.(2003)]{ivan03}
Ivanov, A. A. et al. 2003, Nucl. Phys. B (Proc.Suppl.), 122, 226

\bibitem[Hillas(2006)]{hillas06}
Hillas, A.M. 2006, J. Phys.: Conf. Ser., 47, 168

\bibitem[H\"orandel(2003)]{hoerandel03}
H\"orandel, J.R. 2003, J. Phys. G., 29, 2439

\bibitem[H\"orandel(2005)]{hoerandel05}
H\"orandel, J.R. 2005, astro-ph/0508014

\bibitem[H\"orandel(2007)]{hoerandel07}
H\"orandel, J.R. 2007, astro-ph/0702370

\bibitem[Lucek \& Bell(2000)]{lb00}
Lucek, S. G. \& Bell, A. R. 2000, \mnras, 314, 65

\bibitem[Panov et al.(2006)]{atic2}
Panov, A. D. et al. 2006, astro-ph/0612377

\bibitem[Ptuskin et al.(1997)]{ptuskin97}
Ptuskin, V.S. et al. 1997 \aap, 321, 434

\bibitem[Ptuskin \& Zirakashvili(2003)]{pz03}
Ptuskin, V.S. \& Zirakashvili, V.N. 2003, \aap, 403, 1

\bibitem[Ptuskin \& Zirakashvili(2005)]{pz05}
Ptuskin, V.S. \& Zirakashvili, V.N. 2005, \aap, 429, 755

\bibitem[Ptuskin et al.(2006)]{ptuskin06}
Ptuskin, V.S. et al. 2006, \apj, 642, 902

\bibitem[Takeda et al.(2003)]{takeda03}
Takeda, M. et al. 2003, Astropart. Phys. 19, 447

\bibitem[V\"olk et al.(2003)]{vbk03}
V\"olk, H.J., Berezhko, E. G., \& Ksenofontov, L. T. 2003, \aap, 409, 563

\bibitem[V\"olk(2003)]{vlk03}
V\"olk, H.J. 2003, 
Proc. 28th ICRC,
Invited papers, Vol.8, p.29

\bibitem[V\"olk \& Zirakashvili(2004)]{vz04}
V\"olk, H.J. \& Zirakashvili, V.N. 2004, \aap, 417, 807

\bibitem[V\"olk et al.(2005)]{vbk05}
V\"olk, H. J., Berezhko, E. G., \& Ksenofontov, L. T. 2005, \aap, 433, 229

\end{thebibliography}
\end{document}